\documentclass[a4paper,twoside,notoc,12pt]{JHEP}
\usepackage{amssymb}

\newcommand{\beq}{\begin{equation}}
\newcommand{\eeq}{\end{equation}}
\newcommand{\beqa}{\begin{eqnarray}}
\newcommand{\eeqa}{\end{eqnarray}}
\newcommand{\nn}{\nonumber}

\newcommand{\spa}{\ \ \ }
\newcommand{\eqref}[1]{(\ref{#1})}
\newcommand{\tr}{\mathop{{\rm Tr}}}

\title{The Discrete Bound State Spectrum of the Rotating D0-brane System}
\author{Konstantin G. Savvidy\\
        The Niels Bohr Institute\\
        Blegdamsvej 17, DK-2100 Copenhagen \O, Denmark\\
        E-mail: \email{savvidis@nbi.dk}
        }

\preprint{\hepth{0004113}}

\abstract{
In this note we obtain the discrete spectrum of the rotating 
ellipsoidal membrane, the solution to classical equations of motion in
the matrix mechanics of N D0-branes. This solution has the 
interpretation of a closed D2-brane with the D0-branes bound to
its surface. 
The semi-classical quantization is performed on the rotational modes
with the result that both radii and energy are quantized.  We also
argue that the quantum mechanics of this system is well defined, with
a unique ground state of positive energy in each sector with a
non-zero angular momentum.  The scaling of the size and energy of
these states allows us to identify our rotating brane excited states
with the previously conjectured resonances in the scattering of
D0-branes. }


\begin{document}
 
\section{Introduction} 
\label{intro} 
In this note we elaborate on our previous suggestion in
\cite{Harmark:2000na} , and obtain the discrete spectrum of the
rotating ellipsoidal membrane, the solution to classical equations of
motion in the matrix mechanics of N D0-branes. This solution has the
interpretation of a closed D2-brane with the D0-branes bound to its
surface. We call it closed because it has the shape of an ellipsoid,
which is in addition rotating transversely.  Presumably such solution
exists even in the decompactification limit, where it corresponds to a
state of the fundamental eleven-dimensional M2 brane.  In the latter
case the D0-brane charge in 10D appears just as a result of
Kaluza-Klein reduction.  Such a configuration would play a significant
role in rest-frame M-theory of Witten \cite{Witten:1995ex}.

The main purpose of this note is to quantize
the rotational modes of the system, and we find the exact expression for the energy 
in terms of the  angular momentum  quantum numbers.
We imposed the condition of 
semi-classical or Bohr-Sommerfeld quantization on the action-angle variables.
We obtain the spectrum, {\it i.e.} the discrete energy levels of the system, by
expressing the energy in terms of positive-integer quantum numbers.
In radial coordinates, the quantization proceeds in two steps:
first, angular action-angle variable is quantized by introducing
the angular momentum quantum number $l$; and second, the radial motion
is considered, producing the radial quantum number $n_r$.
An essential difference with the well-understood examples of the 
hydrogen atom and the spatial harmonic oscillator arises: 
in both cases the result for the quantized energy levels 
turns out to depend only on the principal quantum number $n$,
which is the sum of the two $n = n_r +l$.
Moreover this procedure gives the exact answer.
In our system, the first step is almost guaranteed to give
the exact result because angular momentum operator is sure to
appear in the full $QM$. Thus we expect our result to be most reliable
in the case when $n_r=0$, and we call this the "leading Regge" trajectory
because it is the lowest mass state for a given angular momentum.
The correction to the energy due to radial motion, as a function
of the radial quantum number $n_r$ is also sketched out in Section~\ref{sec:qm}.

The D2-brane, being a three-dimensional object, may have at most three
independent components of the angular momentum tensor $M_{ij}$.
Choosing the coordinate system such that this tensor is in a canonical
block-diagonal form, the non-zero components of $M_{ij}$ are
$M_{12}=-M_{21}$, $M_{34}=-M_{43}$ and $M_{56}=-M_{65}$.  We show in
Section \ref{sec:qm} that the system can be reduced, by using angular
momentum conservation, to the following three dimensional radial
potential in cylindrical coordinates
\begin{equation}
    V(\vec{r}) = \frac{{\hat{l}_1}^2}{r_1^2} + \frac{{{\hat{l}_2}}^2}{r_2^2} + 
    \frac{{{\hat{l}_3}}^2}{r_3^2} +
    {\alpha^2}\,  ( r_1^2 r_2^2  + r_2^2 r_3^2 + r_1^2 r_3^2 )
    \label{eq:rpot}
\end{equation}
The extensively studied YM mechanics \cite{Baseian:1979zx,Savvidy:1985gi}
is usually considered with  $ {\mathbf\mathcal l_1} = l_2 = l_3 = 0 $, and
a non-compact moduli space where $V(\vec{r}) =0$, that leads 
\cite{deWit:1988ig,deWit:1989ct} to the 
spreading of the ``particle'' wave-function along the axis and into the ``channel''%
\footnote{The widely held belief that the YM quantum mechanics,
the D0-brane system, the supermembrane and consequently M-theory
all have a continuous spectrum is largely based on this observation.}.
In the general case, with angular momentum turned on, the centrifugal
potential repels the ``particle'' from the ``channel''.
This must lead to a stable ground state, localized at the absolute minimum
of the potential $V$ : the point where 
$ \partial V / \partial\vec{r}=0$,
\begin{eqnarray}
    {{\hat{l}_1}}^2 = \alpha^2 r_1^4 ( r_2^2 + r_3^2) \nn\\
    {{\hat{l}_2}}^2 = \alpha^2 r_2^4 ( r_1^2 + r_3^2) \nn\\
    {{\hat{l}_3}}^2 = \alpha^2 r_3^4 ( r_1^2 + r_2^2) 
    \label{eq:eqn}
\end{eqnarray}
which is exactly the relation that we found in \cite{Harmark:2000na}
for our classical solution.  By solving the above equations for $r_i$,
and then evaluating the energy we get in Section \ref{sec:quant} the
following result for the total energy of the system as a function of
${{{l}_i}}^2$, in a certain approximation,
\begin{equation}
{l_s^3\, N^3\over g_s}E^3= 
    {l_1}^2 \,{l_2}^2 +{l_2}^2 \,{l_3}^2 
    +{l_1}^2 \,{l_3}^2 +
 {5\over 12} \, ({l_1}^2 +{l_2}^2 +{l_3}^2 )^2
    \label{eq:E3intro}
\end{equation}
This result  is subject to the restriction to states with vanishing
radial quantum number $n_r^i=0$, {\it i.e.} for the
leading Regge trajectory.

The exact result was also obtained, and we refer the reader to
the  Section \ref{sec:quant} for the exact formula,
and the conditions for using the approximating formula.

It is our expectation that the whole construction of the solution,
and the quantized energy levels can be carried over to the M2-brane
in eleven dimensions. One possible way to do this is to retell the
story word-by-word in the BFSS \cite{Banks:1997vh}
infinite-momentum frame. Our hope is however
to apply these ideas to a rest frame fundamental M2-brane. 
The connection between M-theory with Planck mass 
$M_p^{-1}= g_s^{1/3}\,l_s$ on a circle
of radius $R_{11}=g_s\,l_s$ and type IIA theory with $g_s$ and 
$\alpha^{\prime}=1/(2\pi l_s^2)$
tells us that had M-theory been one with a continuous spectrum,
the only massive states in IIA would be the KK and wrapping 
modes on the circle, corresponding to the D0-brane and the 
fundamental string respectively.
This may or may not be the case, but it is not unnatural to
suggest that uncompactified M-theory should have an infinite
tower of discrete massive states of its fundamental object,
the M-brane.

\section{The System} 
\label{setup} 

The effective action of $N$ D0-branes for weak and slowly varying fields 
is the non-abelian SU(N) Yang-Mills action
plus the Chern-Simons action (for the bosonic part). 
For weak fields the  action is gotten by dimensionally reducing 
the action of 9+1 dimensional $U(N)$ Super Yang-Mills theory
to 0+1 dimensions \cite{Witten:1996im}. 
Up to a constant term it is
\begin{equation}
\label{BIaction}
S_{} = - T_0 (2\pi l_s^2)^2 \int dt \, 
 \tr \left(\frac{1}{4} F_{\mu \nu} F^{\mu \nu} \right)~~,
\end{equation}
where $F_{\mu \nu}$ is the non-abelian $U(N)$ field strength
in the adjoint representation and \( T_0 = ( g_s l_s )^{-1} \)
is the D0-brane mass.
To write this action in terms of coordinate matrices \( X^i \),
one has to use the dictionary 
\begin{equation}
A_i = \frac{1}{2\pi l_s^2} X^i  ,\spa
F_{0i} = \frac{1}{2\pi l_s^2} \dot{X}^i  ,\spa
F_{ij} = \frac{-i}{(2\pi l_s^2)^2} \, [X^i,X^j]
\label{eq:dict}
\end{equation}
with \( i,j = 1,2,...,9 \), giving 
\begin{equation}
\label{ham}
S_{} = T_0 \int dt \, \tr \left(
\frac{1}{2} \dot{X}^i \dot{X}^i
+ \frac{1}{4} \frac{1}{(2\pi l_s^2)^2 } [X^i,X^j][X^i,X^j] \right)
\end{equation}
To derive this it is necessary to gauge the $A_0$ potential  away,
which is possible for a non-compact time. The Gauss constraint
\begin{equation}
\label{eq:gauss}
[\dot{X}^i,X^i] = 0
\end{equation}
persists, and the above action 
should be taken together with it \cite{Matinian:1981dj}.

The rest of this section is devoted to the discussion
of the relation between currents constructed out of YM fields,
and the corresponding membrane quantities. 

Recently it has become clear \cite{Taylor:1999pr,Myers:1999ps} 
that even though the D0-brane 
world-volume is only one-dimensional, a multiple D0-brane system 
can also couple to  the brane charges of higher dimension.
The Chern-Simons action derived in \cite{Taylor:1999gq,Taylor:1999pr,Myers:1999ps}
for the coupling of $N$ D0-branes to bulk RR $C^{(1)}$ and $C^{(3)}$ fields is 
\begin{equation}
\label{CSaction}
S_{CS} = T_0 \int dt \tr \left( C_0 + C_i \dot{X}^i 
+ \frac{1}{2\pi l_s^2 } \left[ C_{0ij}[X^i,X^j]
+ C_{ijk}[X^i,X^j]\dot{X}^k \right] \right)
\end{equation}
%
This Chern-Simons action not only tells us how $N$ D0-branes move in
the weak background fields of Type IIA supergravity, but also what
higher-form fields the D0-branes produce.  Using this, one can see that
it is possible to build $p$-branes of Type IIA string
theory out of D0-branes
\cite{Taylor:1997dy,Taylor:1999pr,Myers:1999ps}.

The idea that a lower-dimensional object under the influence of
higher-form RR fields may nucleate into spherically or
cylindrically wrapped D-brane was proposed by Emparan \cite{Emparan:1998rt}
for the case of fundamental string. Before that, Callan and
Maldacena \cite{Callan:1997kz} constructed a D- or F-string as a BI soliton solution
on the D-brane where the attached string appears as a spike on the brane.

In general the D2-brane couples to the bulk RR field through the
well-known CS coupling
\begin{equation}
  \label{eqWZ}
  S_{CS} = \int C^{3} = \int C_{\mu\nu\rho}J^{\mu\nu\rho} d^3 \sigma
\end{equation}
where $J$ is a three form RR current
\begin{equation}
  \label{eq:RR}
  J^{\mu\nu\rho} = \epsilon^{\alpha\beta\gamma}~%
  \partial_\alpha X^\mu  \partial_\beta X^\nu \partial_\gamma X^\rho~~,
\end{equation}
or in form notation
\begin{equation}
 \vphantom{J}^{*}J_{(3)}^{\mu \nu \rho} 
= dX^{\mu} \wedge dX^\nu \wedge dX^{\rho}~~.
\end{equation}
One can introduce a charge corresponding to this current,
such that it is a world-volume 2-form even though it has three
space-time indices, same as the current
\begin{eqnarray}
  \label{eq:charge}
  Q^{\mu\nu\rho}_{\beta\gamma} &=&  
   X^{[\mu}  \partial_\beta X^\nu  \partial_\gamma X^{\rho]}~~,~~{\rm so~that~~}\nn\\
  J^{\mu\nu\rho}_{\alpha\beta\gamma} &=& 
3 \, \partial_{[\alpha} Q^{\mu\nu\rho}_{\beta\gamma]}~~,
\end{eqnarray}
or in form notation
\begin{eqnarray}
  \label{eq:dQ}
  Q_{(2)}^{\mu \nu \rho} 
&=& \left(dX^{[\mu} \wedge dX^\nu \right) X^{\rho]}~~,~~{\rm such~that}\nn \\
  dQ_{(2)}^{\mu \nu \rho} &=& J_{(3)}^{\mu \nu\rho}~~,
\end{eqnarray}
where the exterior derivative is taken with respect to world-volume
indices.  This charge first appeared in the theory of bosonic relativistic membrane
\cite{Biran:1987ae}.  There it was interpreted as a topological charge
of the membrane. The connection between RR charges and D-branes  
was discovered by Polchinski \cite{Polchinski:1995mt}.

Instead of (\ref{eq:RR}) one can represent the current as
a Poisson bracket with respect to the spatial world-volume
coordinates. For a static membrane, the non-zero components
of \( J^{\mu\nu\rho} \) are
\begin{equation}
  \label{eq:Poisson}
  J^{0ij}= \left\{ X^i,X^j \right\}~~,
\end{equation}
For a moving membrane the completely spatial components also appear. 
A convenient generalization is, in the static gauge
\begin{equation}
  J^{ijk} = \dot{X}^i \left\{ X^j,X^k \right\}~~.
\end{equation}

The above discussion is for the ordinary D2-brane, and the
coupling of the D0-brane matrix-mechanical system to the 
$C^{(3)}$ field is completely analogous 
\begin{equation}
  \label{eq:WZmyers}
  S_{CS} = \frac{T_0}{2\pi l_s^2 } \int dt~ \tr
  \left( C_{0ij}[X^i,X^j]+ C_{ijk}[X^i,X^j]\dot{X}^k \right)=
  \int C \cdot J~ dt 
\end{equation}
The trace of the first term is zero identically, reflecting the fact
that the total bare RR charge of the object that we are considering is
zero, due to cancellation of the pieces with opposite orientation on
the 2-sphere.
It was proposed in \cite{Taylor:1999gq,Myers:1999ps}, that 
one needs to expand the expression in powers of $X$, to effectively obtain 
the multipole expansion of~$C^{(3)}$. Then, it is possible to
integrate by parts, to get instead the coupling of the field strength
to what we shall call D2-brane dipole moment $Q$:
\begin{equation}
\label{eq:SC}
  S_{CS} =\frac{ T_0}{2\pi l_s^2 } \int dt~
   F_{0ijk} \tr \left[ X^i, X^j \right] X^k = 
   \int F \cdot Q~ dt
\end{equation}

\section{The Rotating Membrane Solution}
\label{sec:Rot}

Next, we review  a new kind of rotating solution to the
system of $N$ D0-branes, which was constructed in 
our previous work \cite{Harmark:2000na}.
There we also showed that it is a stable object, 
in the sence of stability under small perturbations of initial conditions.
Moreover it has  interesting physical properties due to its
dynamical nature, like the radiation of various SUGRA fields.
The basic idea in the construction 
is that the attractive force of tension
should be cancelled by the centrifugal repulsion force.
The motion is at all times transverse, and in no way
can be gauged away by coordinate reparametrization invariance
on the membrane. We refer the reader to the original paper
for more detail.

The D2-brane, being a three-dimensional object,  may have at most
3 independent components of the angular momentum tensor $M_{ij}$.
Choosing the coordinate system such that this tensor is in a 
canonical block-diagonal from, the non-zero components of $M_{ij}$
are $M_{12}=-M_{21}$, $M_{34}=-M_{43}$ and $M_{56}=-M_{65}$.
This argument is the same as for  a particle in a central field, which
always moves in a plane, and so has only one non-zero component of angular momentum.

Therefore, the freedom to choose the 
coordinate system, combined with the global SU(N) rotations, 
means that the very special ansatz we 
considered, may in fact describe the 
dynamics under much more 
general initial conditions.  Hopefully, it will capture all the essential 
features of stationary states of the closed membrane.

We now review the construction of the  rotating ellipsoidal membrane,
viewed as a non-commutative collection of moving D0-branes, 
described in the non-relativistic limit by YM classical mechanics.
For that we need to take the basic configuration \cite{Kabat:1997im}
of the non-commutative fuzzy sphere
in the 135 directions, and set it to rotate in the transverse space
along three different axis, \textit{i.e.} in the 12, 34 and 56 planes.
We thus use a total of 6 space dimensions to embed our D-brane system.
The corresponding ansatz is%
%
\begin{eqnarray}
\label{ansatz6}
&& 
X_1(t) = \frac{2}{\sqrt{N^2-1}}\, \mathbf{T}_1\, r_1(t) \spa,~\spa
X_2(t) = \frac{2}{\sqrt{N^2-1}}\, \mathbf{T}_1\, r_2(t) \spa,
\nn \\ &&
X_3(t) = \frac{2}{\sqrt{N^2-1}}\, \mathbf{T}_2\, r_3(t) \spa,~\spa
X_4(t) = \frac{2}{\sqrt{N^2-1}}\, \mathbf{T}_2\, r_4(t) \spa,
\nn \\ &&
X_5(t) = \frac{2}{\sqrt{N^2-1}}\, \mathbf{T}_3\, r_5(t) \spa,~\spa
X_6(t) = \frac{2}{\sqrt{N^2-1}}\, \mathbf{T}_3\, r_6(t) \spa.
\end{eqnarray}
where the $N \times N$ matrices
$\mathbf{T}_1,\mathbf{T}_2,\mathbf{T}_3$ are the generators of the $N$
dimensional irreducible representation of $SU(2)$, with the algebra
\begin{equation}
\label{SU2alg}
[\mathbf{T}_i,\mathbf{T}_j] = i \, \epsilon_{ijk} \mathbf{T}_k~~.
\end{equation}
In this new ansatz we are using the matrix structure  
such that the coordinate matrices are proportional
to the $SU(2)$ generators in pairs. Simultaneously, the 
Gauss constraint (\ref{eq:gauss}) is identically satisfied.

We interpret this as a rotation
because  one could make a rotation in {\it e.g.} the 12 plane 
which makes one of the  components vanish, say $X_2$, while the  other
one gets a radius $r_1^{\prime} = \sqrt{r_1^2+r_2^2}$. This is 
possible exactly because both are proportional to the same matrix $\mathbf{T}_1$.
The end result is that at any point in time one can choose a coordinate
system in which the object spans only three space dimensions.

Substituting the ansatz into \eqref{ham} gives the Hamiltonian
\begin{eqnarray}
\label{RotHam}
H  =  \frac{NT_0}{3} 
\left( \frac{1}{2} \sum_{i=1}^6 \dot{r}_i^2
+ \frac{\alpha^2}{2}\right. 
\Big[ (r_1^2+r_2^2)\, (r_3^2+r_4^2)~~~~~~~~~~~~~~~~~~~~~~~~~~~~~~~
\nn \\ 
\left.~~~~~~~~~~~~~~~~~~~~~~~~~~
+(r_1^2+r_2^2)\, (r_5^2+r_6^2) 
+ (r_3^2+r_4^2)\, (r_5^2+r_6^2) \Big]\vphantom{\sum_{1}^6}~
\right)~,
\end{eqnarray}
where we have introduced the convenient parameter 
$\alpha = \frac{2}{\sqrt{N^2-1}}\, \frac{1}{2\pi l_s^2}$.

The corresponding equations of motion are
\begin{eqnarray}
\label{eom6}
&&
\ddot{r}_1 = - \alpha^2\,  ( r_3^2+r_4^2+r_5^2+r_6^2 )\, r_1~,\spa
\ddot{r}_2 = - \alpha^2\,  ( r_3^2+r_4^2+r_5^2+r_6^2 )\, r_2~,
\nn \\ && 
\ddot{r}_3 = - \alpha^2\,  ( r_1^2+r_2^2+r_5^2+r_6^2 )\, r_3~,\spa 
\ddot{r}_4 = - \alpha^2\,  ( r_1^2+r_2^2+r_5^2+r_6^2 )\, r_4~, 
\nn \\ &&
\ddot{r}_5 = - \alpha^2\,  ( r_1^2+r_2^2+r_3^2+r_4^2 )\, r_5~,\spa
\ddot{r}_6 = - \alpha^2\,  ( r_1^2+r_2^2+r_3^2+r_4^2 )\, r_6~. 
\end{eqnarray}
We have found the special solution to these equations,
describing a rotating ellipsoidal membrane with 
three distinct principal radii $R_1$, $R_2$ and $R_3$
\begin{eqnarray}
\label{solutions}
&&
r_1(t) = R_1 \cos( \omega_1 t + \phi_1 )~,\ \ 
r_2(t) = R_1 \sin( \omega_1 t + \phi_1 )~,
\nn \\ &&
r_3(t) = R_2 \cos( \omega_2 t + \phi_2 )~,\ \ 
r_4(t) = R_2 \sin( \omega_2 t + \phi_2 )~,
\nn \\ &&
r_5(t) = R_3 \cos( \omega_3 t + \phi_3 )~,\ \ 
r_6(t) = R_3 \sin( \omega_3 t + \phi_3 )~.
\end{eqnarray}
This particular functional form of the solution ensures that the
highly non-linear equations for any of the components $r_i$ are
reduced to a harmonic oscillator.  The solution \eqref{solutions}
keeps $r_1^2+r_2^2=R_1^2$ , $~r_3^2+r_4^2=R_2^2~$ and
$~r_5^2+r_6^2=R_3^2$ fixed
which 
allows us to say that the object described by \eqref{solutions}
rotates in six spatial dimensions as a whole without changing its basic
shape.

Using the equations of motion \eqref{eom6}, the three
angular velocities are determined by the radii, and
do not necessarily have to coincide: 
\begin{equation}
\label{Omega}
\omega_1 = \alpha \sqrt{ R_2^2 + R_3^2 }~,\spa
\omega_2 = \alpha \sqrt{ R_1^2 + R_3^2 }~,\spa
\omega_3 = \alpha \sqrt{ R_1^2 + R_2^2 }~.\spa
\end{equation}
This dependence of the angular frequency on the radii 
is such that the repulsive force of rotation 
has to be balanced with the attractive force of tension
in order for \eqref{solutions} to be a solution. 
Thus the radii $R_1$, $R_2$ and $R_3$ parameterize \eqref{solutions} 
along with the three phases
$\phi_i$, to produce altogether a six parameter family of solutions.

Next, we evaluate the energy,
\begin{equation}
E = \frac{NT_0}{4} \left( \omega_1^2 R_1^2 +
    \omega_2^2 R_2^2 + \omega_3^2 R_3^2 \right)~~.
\label{enEv}
\end{equation}

In order to exhibit the properties of the solution \eqref{solutions}
we compute the components of the angular momentum 
\begin{equation}
  \label{Mij}
  M_{ij}= \tr \Big[ X^i \Pi^j -X^j \Pi^i \Big]~~,~~{\rm where}~~~\Pi^i = T_0 \dot{X}^i~~.
\end{equation}
As expected, the non-zero components are time-independent and
equal to
\begin{equation}
\label{Mexp}
M_{12} = \frac{1}{3}N T_0 \omega_1 R_1^2 ~~,~~~
M_{34} = \frac{1}{3}N T_0 \omega_2 R_2^2 ~~,~~~
M_{56} = \frac{1}{3}N T_0 \omega_3 R_3^2 ~~.
\end{equation}
The angular momenta $M_{12}$, $M_{34}$ and $M_{56}$ correspond to rotations in
the 12, 34 and 56 planes respectively. Their values \eqref{Mexp} fit with the
interpretation of the solution as $N$ D0-branes rotating as an ellipsoidal
membrane in that they are time-independent due to conservation law and
proportional to \( N T_0 \omega_i R_i^2 \).

\section{Quantum Mechanics}
\label{sec:qm}

The Hamiltonian \eqref{RotHam} can be readily used to consider the 
corresponding quantum mechanical problem. However,
a more thorough investigation of the procedure of reduction
to radial coordinates (see below) is necessary in order to 
completely describe the D0-branes in 9+1 dimensions.

The picture is that of a particle in a six-dimensional space, and with a positive
definite potential energy. It does not have an isolated minimum of the energy,
but has instead a moduli space which is non-compact. The wave function spreads in
those directions in the moduli space, the result being that there is no
normalizable ground state with a positive energy. This does not mean that the
theory is sick, but rather that it has a superimposed continuum spectrum, in
addition to the discrete. In fact, the existence of this continuum is required if
the system is to reduce to supergravity in the low-energy limit.

The conservation of angular momentum allows us to reduce the
problem to a three-dimensional one in radial coordinates:
\begin{eqnarray}
r_1 = \rho_1 \cos\phi_1 ~,& ~r_2 = \rho_1 \sin\phi_1 ~,\nn\\
r_3 = \rho_2 \cos\phi_2 ~,& ~r_4 = \rho_2 \sin\phi_2 ~,\nn\\
r_5 = \rho_3 \cos\phi_3 ~,& ~r_6 = \rho_3 \sin\phi_3 ~.
\end{eqnarray}
In these radial coordinates, the six-dimensional Laplacian
can be broken up into radial and angular parts, with the angular
part being computable from \eqref{Mexp}
\begin{equation}
\label{eq:schroe}
\left[ -\left(  {\partial^2 \over \partial\rho_1^2} +  
{\partial^2 \over \partial\rho^2_2} +
{\partial^2 \over \partial\rho^2_3}  \right) + 
{\alpha^2}  ( \rho_1^2 \rho_2^2  + \rho_2^2 \rho_3^2 + \rho_1^2 \rho_3^2 ) +
\frac{{\hat{l}_1}^2}{\rho_1^2} + \frac{{\hat{l}_2}^2}{\rho_2^2} + 
    \frac{{\hat{l}_3}^2}{\rho_3^2} \right] \Psi = 2 \, E \Psi
\end{equation}
where we have identified the classical values of the angular
momentum components $M_{ij}$ with the corresponding 
quantum mechanical operator $\hat{l}$.
In the non-degenerate case, when at least two of the $l$'s are
non-zero, the above potential has an isolated global minimum at
\begin{eqnarray}
    {\hat{l}_1}^2 = \alpha^2 \rho_1^4 ( \rho_2^2 + \rho_3^2) \nn\\
    {\hat{l}_2}^2 = \alpha^2 \rho_2^4 ( \rho_1^2 + \rho_3^2) \nn\\
    {\hat{l}_3}^2 = \alpha^2 \rho_3^4 ( \rho_1^2 + \rho_2^2) 
    \label{eq:eq}
\end{eqnarray}
The comparison with \eqref{Mexp} and \eqref{Omega} shows that our 
classical solution (\ref{solutions})
indeed is the one at the stable minimum of the potential. From here
it is now clear that perturbations away from the equilibrium point
are stable, and cause almost-harmonic oscillations around it. These can
be analyzed by considering the tensor of the second order
derivatives of the potential in \eqref{eq:schroe} 
\begin{eqnarray}
   \left( 
 \begin{array}{c}
 6\, {l_1^2 \over \rho_1^4} + \alpha^2\, ( \rho_2^2+\rho_3^2) \spa\spa 2 \alpha^2 \rho_1\rho_2 \spa\spa 2 \alpha^2 \rho_1\rho_3\\
2 \alpha^2 \rho_1\rho_2 \spa\spa  6\, {l_2^2 \over \rho_2^4} + \alpha^2\, ( \rho_1^2+\rho_3^2) \spa\spa 2 \alpha^2 \rho_2\rho_3\\
2 \alpha^2 \rho_1\rho_3 \spa\spa 2 \alpha^2 \rho_2\rho_3 \spa\spa 6\, {l_1^2 \over \rho_3^4} + \alpha^2\, ( \rho_1^2+\rho_2^2)
  \end{array} \right)~~,
    \label{eq:matrix}
\end{eqnarray}
At the equilibrium point of (\ref{eq:eq}) this is equal to
\begin{eqnarray}
  4 \alpha^2 \left( 
 \begin{array}{c}
  2(\rho_2^2+\rho_3^2)\spa\spa \rho_1\rho_2 \spa\spa \rho_1\rho_3\\
  \rho_1\rho_2 \spa\spa 2 \, ( \rho_1^2+\rho_3^2) \spa\spa \rho_2\rho_3\\
  \rho_1\rho_3 \spa\spa  \rho_2\rho_3 \spa\spa 2\, ( \rho_1^2+\rho_2^2)
  \end{array} \right)~~,
    \label{eq:simpmatrix}
\end{eqnarray}
By diagonalizing this matrix, we obtain the frequencies of the
oscillations around the stable minimum, as well as the energies of the 
corresponding harmonic oscillator quantum mechanical states. 
The eigenvalues are everywhere positive:
\begin{eqnarray}
\lambda_1 &=& 2\, \left( {{ \rho_1}}^{2}+{{ \rho_2}}^{2}+{{ 
\rho_3}}^{2} \right) \nn\\
\lambda_{2,3} &=& {{ \rho_1}}^{2}+{{ \rho_2}}^{2}+{{ \rho_3}}^{2}\pm
{1 \over \sqrt 2} \, \sqrt{
({{ \rho_1}}^{2}-{{ \rho_2}}^{2})^2 + ({{ \rho_2}}^{2}-{{ \rho_3}}^{2})^2 +
({{ \rho_3}}^{2}-{{ \rho_1}}^{2})^2 } \nn
\end{eqnarray}
This allows us to  approximately evaluate the energies of the states that are
close, but not lying on the leading Regge trajectory. The simple formula below
gives a correction to the energy (\ref{enEv}) due to the radial oscillations, in terms of the
three radial quantum numbers $n_r$ ,
\begin{equation}
  \label{eq:radialN}
  \delta E = \sqrt{\lambda_1} \, n_r^1 + \sqrt{\lambda_2} \, n_r^2 + \sqrt{\lambda_3 }\, n_r^3 ~~.
\end{equation}

\section{Energy Quantization}
\label{sec:quant}

We have previously shown \cite{Harmark:2000na} 
that the rotating ellipsoidal membrane solution is 
classically stable. 
However the system constantly loses energy
due to the semi-classical radiation of various supergravity waves.
As a side, we estimate the life-time of the excited states:
it is the ratio of the kinetic energy of the system to
the rate of radiation per unit time,
\begin{equation}
  \label{eq:life}
  \tau_{1/2} = \frac{E}{P} \sim 
  \frac{N\, T_0 \,\omega^2 R^2}{\,\kappa^2 N^2\, T_0^2\, \omega^{12} R^4\, } =
  \left( \kappa^2 N\, T_0\, \omega^{10} R^2 \right)^{-1}~~.
\end{equation}
The substitution of the various quantities, including the 
characteristic size $R$ of the system, which will turn out to be
the eleven-dimensional Planck length $g_s^{1/3} l_s$ yields
\begin{equation}
  \label{eq:life1}
  \tau_{1/2} \sim {1\over g_s^5 \, l_s}\, N^9 ~~.
\end{equation}
Thus the states are not stable in the absolute sence, but nevertheless are 
long-lived, compared to the string scale. We emphasize again, that the intrinsic
dynamics of the system dictates a zero width. 

What will happen to the 
rotating ellipsoidal membrane  after all
available kinetic energy has been radiated?
The radiated quanta carry away  angular momentum as well,
eventually going into $l=0$ state. As discussed above,
the sector with vanishing angular momentum has a continuous spectrum,
so the system will, most likely, decay into $N$ free D0-branes.

However, for non-vanishing angular momentum,
the Bohr-Sommerfeld
quantization rule tells us that the angular momentum should be
quantized in units of the Planck constant.

We use a scheme, due to Ehrenfest, in which an adiabatic invariant
of the form $ \int p_i dx_i $, which turns out to be equal to
the angular momentum, is quantized in units of the Planck constant,%
\footnote{The Planck constant $\hbar$ has been set to unity throughout.}
\begin{eqnarray}
\label{Mquant}
{1\over 2\pi} I_1 = {1\over 2\pi} \tr \int \Pi_1 dX_1 + \Pi_2 dX_2=
M_{12} = \frac{1}{3}N T_0 \omega_1 R_1^2 =  l_1 ~~,\nn\\
{1\over 2\pi} I_2 = {1\over 2\pi} \tr \int \Pi_3 dX_3 + \Pi_4 dX_4=
M_{34} = \frac{1}{3}N T_0 \omega_2 R_2^2 =  l_2 ~~,\nn\\
{1\over 2\pi} I_3 = {1\over 2\pi} \tr \int \Pi_5 dX_5 + \Pi_6 dX_6=
M_{56} = \frac{1}{3}N T_0 \omega_3 R_3^2 =  l_3 ~~.
\end{eqnarray}
%
So far we are safe, as the quantization of the angular momentum
as integer values of $l_i$ is expected to be an exact feature
of the complete quantum mechanics. 

We should combine these equations with \eqref{Omega}
and solve for $R_i$. Then reexpress the energy (\ref{enEv}) 
in terms of $l_1$, $l_2$ and $l_3$ only:
\begin{eqnarray}
    {l_s^3\, N\,(N^2-1) \over g_s} \, E^3= 
    l^2_{1}\,l^2_{2}+l^2_{2}\,l^2_{3}+l^2_{1}\,l^2_{3}+~~~~~~~~~~~~~~~~~~~~\nn\\
    ~~~~~~~~\frac{1}{12}\,\left (
    {\zeta} + \frac{{\psi}^{2}}{\zeta}-\psi
    \right ) \left (
    {\zeta} + \frac{{\psi}^{2}}{\zeta}+3\,\psi\right )
  \label{eq:E3}
\end{eqnarray}
where 
\begin{eqnarray}
    \zeta^3 = {54\,\chi-{\psi}^{3}+6\,\sqrt {81\,{\chi}^{2}-3\,\chi\,{
\psi}^{3}}}\nn\\
    \psi\,  = l^2_1+l^2_2+l^2_3 ~~\mbox{and}~~ \chi = l^2_1 l^2_2 l^2_3
    \label{eq:chipsi}
\end{eqnarray}
The uncertainty is, however, in the total angular momentum. 
We have used the naive value, $l^2$, but one might wonder whether
the exact result can be gotten by a different identification, such as
${\hat{l}_i}^2=(l+1)^2$. One argument in favor of this is the WKB
approximation formulae, which dictates adding a unity to the $RHS$
of (\ref{Mquant}). In any case, it is not clear whether the semi-classical 
approximation is  miraculously capable of yielding
the exact result, as is the case in the hydrogen atom.

Nevertheless, the principal message is that the
bound-state spectrum of the D0-brane system is indeed discreet,
and that there is a mass gap in every sector with non-zero 
relative angular momentum.

The result (\ref{eq:E3}) is quite a bit complicated, but we shall see
that it simplifies under certain assumptions. In 
particular, let $n_i$ be large but approximately
equal to each other (meaning their mutual ratios are close to one),
then
\begin{eqnarray}
    {l_s^3\, N\,(N^2-1) \over g_s}
    E^3= 
    l^2_{1}\,l^2_{2}+l^2_{2}\,l^2_{3}+l^2_{1}\,l^2_{3}+
    {5\over 12} \, (l^2_1+l^2_2+l^2_3)^2
  \label{eq:E3simp}
\end{eqnarray}

The opposite limit, that of widely disparate values of $l_i$,
can also be addressed analytically and gives the following result
\begin{equation}
    {l_s^3\, N\,(N^2-1) \over g_s}
    E^3= 
    l^2_{1}\,l^2_{2}+l^2_{2}\,l^2_{3}+l^2_{1}\,l^2_{3}+
    2\, l_1\, l_2 \, l_3 \sqrt{l^2_1+l^2_2+l^2_3}~~.
    \label{eq:E3zero}
\end{equation}

The characteristic scale of $R$, the analog of the Bohr radius, is the
eleven dimensional Planck length $ g_s^{1/3}l_s$.  The mass scale, an
analog of the Rydberg constant,%
\footnote{We point out that the apparent incompatibility of the 
mass scale with the eleven dimensional Planck mass goes away after
the "de-boosting" procedure, along with the $N$ dependence.}
is $g_s^{1/3} l_s^{-1} N^{-1}$. Both are
features that were previously expected \cite{Shenker:1995xq}, 
esp. in the D0-brane scattering problem
\cite{Kabat:1996cu,Douglas:1997yp,Danielsson:1996uw}. We would like to suggest the
identification between our rotating mebrane states and the resonances
conjectured to exist in that context.  However, the important
consequence from perturbative string calculations and YM mechanics
(see also Bachas \cite{Bachas:1996kx}) is
that the D0-brane system may probe important eleven dimensional
physics, even in the non-relativistic limit. Certainly, this
lends credibility to our suggestion that the fundamental M2-brane may
form the same kind of closed ellipsoidal rotating configurations. We
stress that progress is evident in this approach, even without the 
advantages of going over to the infinite-momentum frame M(atrix) theory \cite{Banks:1997vh}.

\section{Acknowledgments}

I would like to thank J. Ambj{\o}rn and C. Callan Jr. for
extensive discussions and useful comments on the draft of the paper.
Also many thanks to  
T. Harmark, D. Kabat, R. Myers, P. Pouliot, W. Taylor IV and Z. Yin.

\addcontentsline{toc}{section}{References}

\bibliographystyle{JHEP}
%

\bibliography{membrane_bib}

\raggedright

\end{document}